\documentclass[conference]{IEEEtran}
\IEEEoverridecommandlockouts
\usepackage{cite}
\usepackage{amsmath,amssymb,amsfonts}
\usepackage{algorithmic}
\usepackage{graphicx}
\usepackage{textcomp}
\def\BibTeX{{\rm B\kern-.05em{\sc i\kern-.025em b}\kern-.08em
    T\kern-.1667em\lower.7ex\hbox{E}\kern-.125emX}}
\begin{document}

\title{Constructing Trustworthy and Safe Communities on a Blockchain-Enabled Social Credits System}


\author{
\IEEEauthorblockN{Ronghua Xu, Xuheng Lin, Qi Dong, Yu Chen}
\IEEEauthorblockA{
Dept. of Electrical and Computer Engineering, Binghamton University, SUNY, Binghamton, NY 13902, USA \\
\{rxu22, xlin84, qdong3, ychen\}@binghamton.edu}
}

\maketitle

\begin{abstract}
The emergence of big data and Artificial Intelligence (AI) technology is reshaping the world. While the technological revolution improves the quality of our life, new concerns are triggered. The superhuman capability enables AI to outperform human workers in many data- and/or computing-intensive tasks. Also, digital superpowers are showing arrogance towards individuals, which erodes the trust foundation of the society. In this position paper, we suggest to construct trustworthy and safe communities based on a BLockchain-Enabled Social credits System (BLESS) that rewards the residents who commit in socially beneficial activities. Human being's true value lies in serving other people. The BLESS system is considered as an efficient approach to promote the value and dignity in efforts focused on enhancing our communities and regulating business and private behaviors. 
The BLESS system leverages the decentralized architecture of the blockchain network, which not only allows grassroots individuals to participate rating process of a social credit system (SCS), but also provides tamper proof of transaction data in the trustless network environment. The anonymity in blockchain records also protects individuals from being targeted in the fight against powerful enterprises. Smart contract enabled authentication and authorization strategy prevents any unauthorized entity from accessing the credit system. The BLESS scheme is promising to offer a secure, transparent and decentralized SCS.  
\end{abstract}



\section{Introduction}
\label{sec:intro}  

This position paper suggests constructing more trustworthy and safer communities based on a BLockchain-Enabled Social credits System (BLESS) inspired on two critical observations. One the one hand, the proliferation of big data driven technologies allows governments and private sectors to provide intelligent services leveraging information obtained by collecting and fusing fragmented public and private data, but the superpower in digital economy sometimes leads to their arrogance towards individuals. Recently we have witnessed multiple tragedies, such as what happened at Uber \cite{obrien2018cnn}, DiDi \cite{wang2018didi} and the false vaccine scandal \cite{selby2018vaccine}. These accidents have stirred anxiety across the society. Because of the high cost to individuals to appeal justice in the fighting against powerful enterprises or large organizations, some dishonest enterprises abuse the victims (and the society) without receiving any penalty. What is even worse, years of unscrupulous behaviors, like environmental disasters, academic dishonesty, and food and drug safety scares, could erode trust in the integrity of government actions among the citizens. A mechanism is desperately needed to help grassroots individuals to help each other and protect themselves in this asymmetrical fight.


On the other hand, Artificial Intelligence (AI) technology is reshaping global communications, commercial activities, and social relations in our communities. The superhuman capability in data processing and pattern recognition enables AI to outperform human workers in many data- and/or computing-intensive tasks. In addition, the AI significantly cuts the cost with very high efficiency for manufacturers and service providers. As AI is starting to take over many jobs across the society, there are concerns about the value of human being \cite{drum2017job}, \cite{Lee2018ai}. However, there is a gap between machine and human being: empathy and love. Human being's true value lies in the time and care one devotes in serving other people, not in the skills that AI will master quickly. Therefore, a new value system is desired that rewards the residents who commit in socially beneficial activities. 

To tackle the challenges in social trust relationship and make members accountable, Social Credit System (SCS) is considered as an efficient approach to shore up trust in the public organizations, like government agencies and enterprises, and regulate business and private behaviors conduct by entities. The SCS allows governments, enterprises and individuals to make critical decisions based on analytic data instead of arbitrary whims. Those who have positive credit evaluation are awarded benefits of conducting the benign behaviors to the society. While those who have transgressed, such as failing to pay tax or getting speed ticket, are subject to penalties, which in turn forbid them from receiving various services, like taking airplane or car rental \cite{liang2018constructing}. 

Although the SCS offers a prospective solution to address issues that are related to the trust relationship among government, enterprises and citizen, it still faces many challenges. Currently, only governments or authorized third-party organizations could combine fragmented public and private data to construct a SCS. However, credit score standard is not transparent and individuals do not participate the credit rating procedure to oversee those agents in power. In addition, the ability of credit-scoring organizations to collect, use and share individual's personal data is not constrained. Thus, it is difficult to prevent user's private data from being abused or misused by the agents without proper regulations. Furthermore, traditional credit systems, from financial scoring systems to extended social credit system, are mainly relying on centralized architecture, which are vulnerable to single point of failure problem and become the performance bottleneck.

Blockchain, which acts as the fundamental protocol of Bitcoin \cite{nakamoto2008bitcoin}, has demonstrated great potential to revolutionize the fundamentals of information technology (IT) due to many attractive features, such as decentralization and transparency. The decentralized architecture of the blockchain network provides a promising solution that unites the strength of the ignored individual underdogs to address existed issues in current social credit systems.

In this position paper, we propose the BLESS system, which is considered as an efficient approach to construct more trustworthy and safer communities, promote the value and dignity in efforts focused on enhancing our communities and regulate business and private behaviors. The decentralized blockchain network enables grassroots individuals to participate credit evaluation process instead of relying on centralized third-party authority to manage credit score data. The score and evaluation procedure are transparent to public and consensus mechanism enforced by blockchain provides tamper proof of transaction data in the trustless network environment. The proposed BLESS solution offers a secure, transparent and decentralized social trust infrastructure. It not only will reduce the cost in defending each person's rights, but also avoid to be targeted by the powerful forces such as big enterprises.

The rest of this paper is organized as follows. Section \ref{sec:back_know} provides the background knowledge. The architecture and rationale of the prototype design for BLESS are presented in Section \ref{sec:bless}, and Section \ref{sec:conclusion} concludes this paper with some discussions.

\section{Background Knowledge}
\label{sec:back_know}
\subsection{Credit Scoring System}

Credit scoring is a statistical method of evaluating the credit risk of loan applications. Using historical data and statistical techniques, credit scoring system tries to predict the probability that a loan applicant or existing borrower will default or become delinquent \cite{mester1997s}.  The "score" that is produced by a credit scoring method can be used by banks or financial institutes to rank their loan applicants or borrowers in terms of risk. Because of the benefits, like reducing the time needed in the loan approval process and improving objectivity in the loan approval process, over the past decades, credit scoring has been widely used for consumer lending, such as credit cards issuing and mortgage lending \cite{mester1997s}. The financial credit scoring systems, which are commonly used by financial institutions in the United States, help financial organizations and individual make decisions based on trust and certificated credit score, and improve accounting and loan-lending activities. 

Thanks to big data technology, many governments are considering to leverage the power of big data to change the way they govern by extending traditional commercial credit systems to other areas of governmental regulation \cite{chorzempa2018china}. As one of the most highlighted cases, China's Social Credit System (SCS) has drawn considerable attention from both academic and public not only because this top-level effort tries to integrate traditional commercial credit rating into social behaviors, but also because it involves the construction of centralized data infrastructures for data collection, mining, and analysis \cite{liang2018constructing}. The SCS is used to monitor four areas for social credit scoring: government affairs, judicial affairs, social activities, and commercial behaviors; and it offers a reward and punishment mechanism \cite{liang2018constructing}. Compared to credit systems in other nations like the United States, China's SCS goes far beyond a financial credit rating and is considered to be a ``broad social credit system'' \cite{liang2018constructing}.

\subsection{Blockchain and Smart Contract}

The blockchain technology, which was initially introduced in 2008, has demonstrated its success in decentralization of digital currency and payment, like bitcoin \cite{nakamoto2008bitcoin}. It is a public ledger that provides a verifiable, append-only chained data structure of transactions. By allowing the data be stored and updated distributively, the blockchain is essentially a decentralized architecture that does not rely on a centralized authority. The transactions are approved by a large amount of distributed nodes called miners and recorded in timestamped blocks, where each block is identified by a cryptographic hash and chained to preceding blocks in a chronological order. Blockchain uses consensus mechanism, which is enforced on miners, to maintain the sanctity of the data recorded on the blocks. Thanks to the ``trustless" proof mechanism running on miners across networks, users can trust the system of the public ledger stored worldwide on many different nodes maintained by ``miner-accountants'' as opposed to having to establish and maintain trust with a transaction counter-party or a third-party intermediary \cite{swan2015blockchain}. Thus, Blockchain is considered an ideal decentralized architecture to ensure distributed  transactions between all participants in a trustless environment.

Emerging from the smart property, smart contract allows users to achieve agreements among parties and supports variety of flexible transaction types through blockchain network. By using cryptographic and security mechanisms, smart contract combines protocols with user interfaces to formalize and secure relationships over computer networks~\cite{szabo1997formalizing}. A smart contract includes a collection of pre-defined instructions and data that have been saved at a specific address of a blockchain as a Merkle hash tree, which is a constructed bottom-to-up binary tree data structure. Through exposing public functions or application binary interfaces (ABIs), a smart contract interacts with users to offer predefined business logic or contract agreement. The blockchain and smart contract enabled security mechanism for applications has been a hot research topic and some efforts have been reported recently, for example, smart surveillance system \cite{nikouei2018real, nagothu2018microservice}, identification authentication \cite{hammi2018bubbles} and access control \cite{xu2018blendcac, xu2018smartcac}. We believe blockchain and smart contract together are promising to provide a solution to enable secured credit rating and score recording in decentralized social credit systems.

\subsection{Time Banking}
As a co-production of social services, Time Banking is a generalized exchange economy not based on money, and values everyone's contribution on the same scale (time expended) \cite{carroll2013co}. Time banking has spread rapidly in recent years. For example, the nonprofit organization, TimeBanks USA \cite{TimeBanksUSA} facilitates 276 time banks in North America through 27,000 members, as well as in other countries. In time banking system, all members' time is treated as equal, which allows value created by service exchanges to remain within the local community. Apart from the obvious benefit of allowing people without money or a job to participate in value creation, a timebank creates opportunities for new relationships to form and strengthens bonds among community members \cite{bellotti2014towards}. Some research shows that more active involvement in community actually leads to higher levels of quality of life \cite{kwon2017community}. 

Time banking is a network, which functions as a platform letting people provide and receive services from each other by donate their time. Generally, it does not involve real money. One simple example is that one person can hire another one for massaging for one hour. Then, the masseur earns one hour credits and next time, the masseur can spend the credits to hire other people to work for him/her for one hour. There are five core values of time banking as listed by Edgar Cahn, the founder of modern time banking \cite{cahn2011time}, \cite{cahn2000no}. 

\begin{itemize}
   \item First, the asset, everyone has something to provide, from washing dishes to taking care of elder people, or even providing companying; 
   \item Second, redefining work, some work cannot be found on the market like revitalizing neighborhoods, or is hard to hire someone who is trust worthy; 
   \item Third, reciprocity, this emphasis providing and receiving among the neighbors and thus helps building strong connections between neighborhoods; 
   \item Fourth, social networks, it will be strengthened by time banking because time banking provides a way that allows people to share their skills among each other; and
   \item Fifth, respect, every transaction in the time banking is based on the agreement and trust. And because blockchain is embedded in time banking thus the trust can be guaranteed. 
\end{itemize}

Time banking is a noble idea and its potential has been recognized by more and more people \cite{yuan2018assessing}. However, it has not brought impacts as significant as its inventor expected because of several weaknesses. Some researchers pointed out that psychologically the metaphor of ``bank'' failed to serve its purpose of promoting the community centered, peer-to-peer service exchange \cite{bellotti2014towards}. Technically speaking, one of the weaknesses is that the security and trust issues are not well addressed. In a community consisting of dozens of residents it is very likely that people does not know each other very well. A person may know the others living in the next building but not the one who lives a block away. Therefore, a trustful relationship could not be built in a community and people normally would not take risk of asking someone who lives a block away to help. The blockchain enabled time banking can tackle this issue perfectly and will be discussed in Section \ref{sec:tb}.

\section{BLESS System Architecture}
\label{sec:bless}

In this section, a system level illustration of the BLESS architecture is presented. As a decentralized SCS, the BLESS system tracks and manages the credit scores of the residents of a community, including individuals, enterprises, organizations and government agencies. It encourages honest residents to participate in public affairs by awarding credits to residents who commit in social beneficial activities and reducing the credits of ones whose behavior hurt the harmony or degrade the life quality of the community. Specifically the BLESS is proposed to serve two main purposes: (1) promoting the value and dignity in jobs serving our fellow residents; and (2) protecting grassroots individuals from being targeted when they speak out in certain public affairs.  

In addition, time banking is selected as an application to develop a proof-of-concept prototype of a BLESS system, although a time banking does not consist of all components of a complete BLESS system. A mobile app will be built and testes by student volunteers using it as a case study, through which some preliminary data will be collected for more insights about the characteristics of the system.

\begin{figure} [t]
\begin{center}
\begin{tabular}{c}
\includegraphics[height=10.6cm]{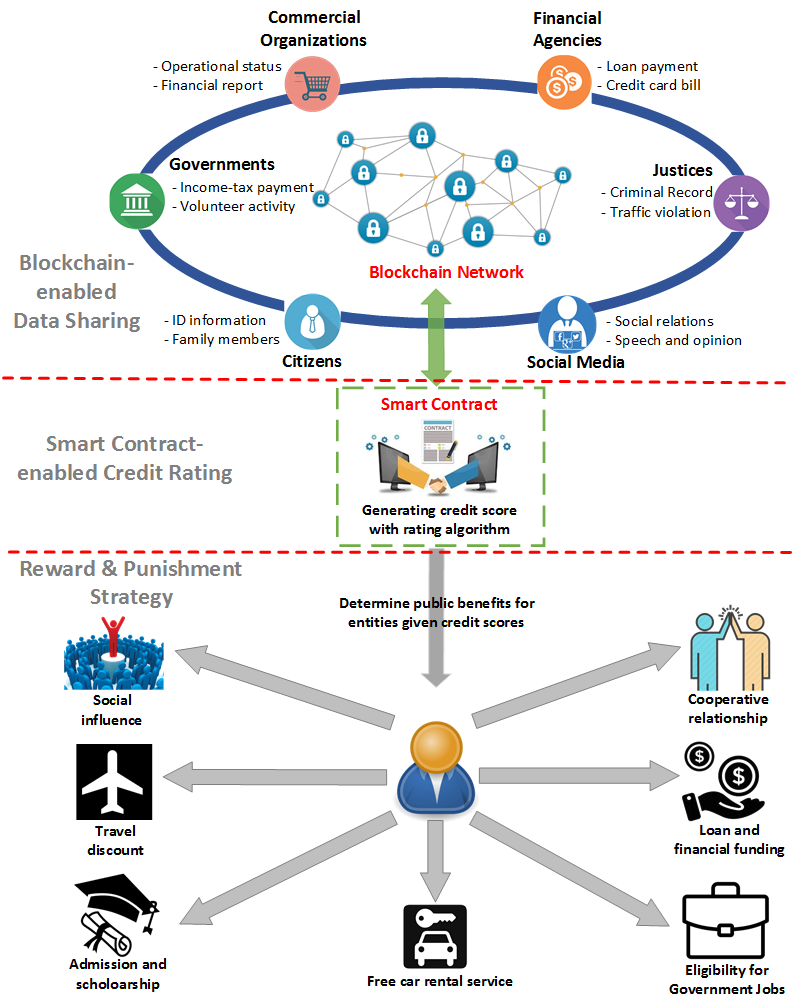}
\end{tabular}
\end{center}
\caption[example] {\label{fig:1-SCS} Illustration of the BLESS System Architecture.}
\vspace{-10pt}
\end{figure}

\subsection{Main Functional Components}

The BLESS System is essentially a decentralized social trust infrastructure including data collection, aggregation, mining and analysis by integrating fragment public and private data resources in the society. Figure \ref{fig:1-SCS} illustrates the architecture. In the BLESS system, various entities in a community, like government agencies, enterprises and individuals, could participate the network and cooperate with each other to construct the decentralized SCS. The rewarding and punishing mechanism based on credit scores could effectively prevent entities who have trust-breaking behaviors records from being influential in the community. 
As shown in Fig. \ref{fig:1-SCS}, the comprehensive BLESS system consists of three components: Blockchain-enabled data sharing, smart contract enabled credit score rating and the rewarding and punishing strategy.

\begin{enumerate}
\item \emph{Blockchain-enabled Data Sharing}: 
In the BLESS system, the data collection function is carried out via multiple stakeholder mechanism, and entities only gather data in which they are concerning and interested. For example, the government agencies have income-tax payment records, while financial agencies would collect financial data, like loan repayment and credit card bill. The distributed entities could use the blockchain network to securely integrate separate data in a trustless network environment. The entity who has been authenticated and authorized an access to system is allowed to submit data transactions to the blockchain network. Thanks to the consensus mechanism enforced by the blockchain, only the confirmed data and transactions can be recorded in a new block and appended to the main chain. Finally, data collected by disparate stakeholders are shared and integrated via the blockchain for social credit rating.

\item \emph{Smart Contract Enabled Credit Score Rating}: 
The credit score rating algorithms are encapsulated and transcoded into smart contracts, which are deployed on the blockchain network by the certificated credit rating entities. After a smart contract has been deployed successfully on the blockchain network, it becomes visible to all the users on the network owing to the transparency and publicity properties of the blockchain. Thus, any participant in the blockchain network can access the transactions and smart contracts recorded in the chain to oversee the rating process and verify credit scores. Because of the cryptographic and security mechanisms provided by the blockchain network, smart contracts can secure any algorithmically specifiable protocols and relationships from malicious interference by third parties in the trustless network environment. After synchronizing the blochchain data, every participant could access all the transactions and the recent states of each smart contract by referring local chain data. Each node interacts with the smart contract through the provided contract address and the remote procedure call (RPC) interface.

\item \emph{Rewarding and Punishing Strategy}: 
It is the kernel of the the BLESS system to serve its purposes. The first one is to promote the value and dignity in jobs serving our fellow residents. The residents in a community can promote their credit scores by either serving the community, for example, take care of senior citizens or children, volunteer to teach courses in community, serve as volunteer fire fighter, etc, or speaking out for social justice, for example, express their support or against in certain public events. In some scenarios, particularly in case when grassroots individuals fight for their right against powerful enterprises, public media, or government agencies, the BLESS system allows them to stand out and other community members can support them without worrying to be targeted. Of course, the expressing of opinion is not for free, certain amount of credit points would be charged. When the investigation is finished, the system will pay back those who spoke honestly by awarding them credit points more than they paid; those who gave false statements will lose the credit they paid. Consequently, the more the honest residents or enterprises contribute the the community, the higher their credit scores are. Those whose behaviors are not accepted by the community will eventually lose their impacts to the public affairs, or maybe even worse, to be denied to some public benefits.

\end{enumerate}

\begin{figure} [t]
\begin{center}
\begin{tabular}{c}
\includegraphics[height=7.5cm]{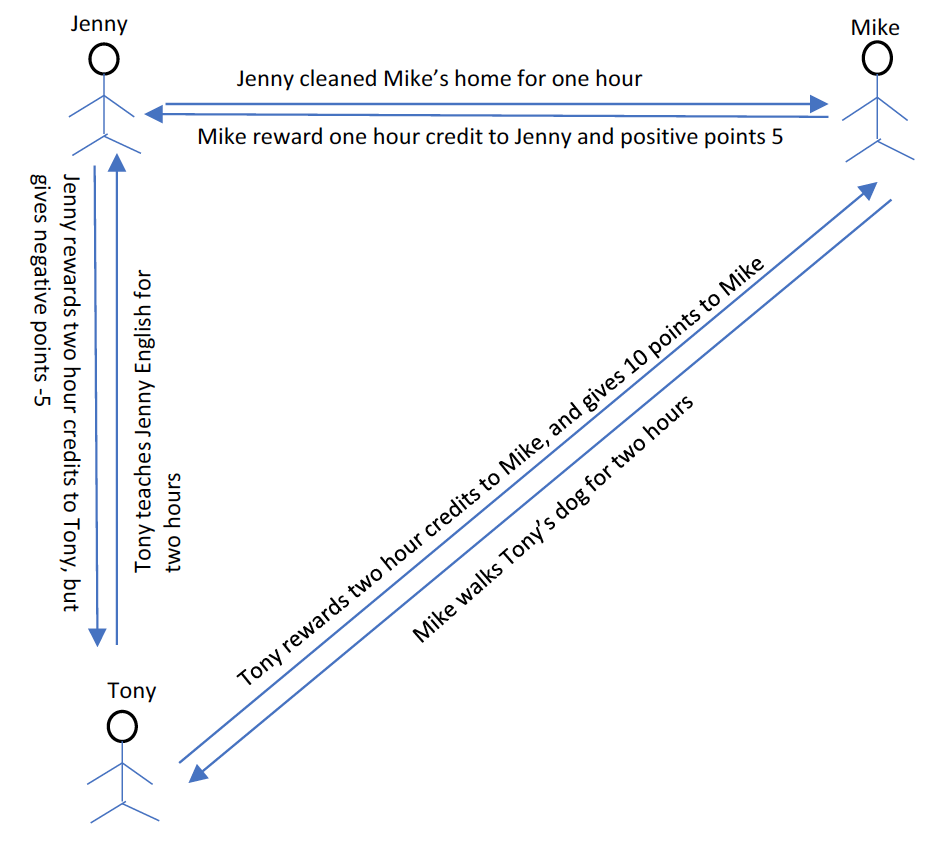}
\end{tabular}
\end{center}
\caption[example] {\label{fig:2-time banking} Illustration of time banking system.}
\vspace{-10pt}
\end{figure}

\subsection{A Blockchain Enabled Time Banking}
\label{sec:tb}

In the preliminary design of BLESS system, time banking is selected as the case to evaluate the feasibility. The blockchain is able to address the weaknesses in the current time banking from several aspects. Different from traditional web application that relies on a centralized server. The blockchain enabled time banking uses Ethereum platform to achieve its goal of decentralization. Thus, any services provided in the time banking will be recorded and monitored by each node and this ensures the security and transparency. Moreover, a smart contract will be defined in the time banking to guarantee that the transaction between each user meets the security level. For example, before two users make the agreement on a house clean service, their information will be passed into the smart contract. The information may include their credit score, service history, conviction record, and so on. The smart contract is responsible to make sure that this information meets the criterion in which users can trust. 

In a time banking implementation on top of the blockchain, the credit points are given to each user. Figure \ref{fig:2-time banking} illustrates how the time banking system works with an example. If Jenny spent one hour to clean Mike's home, then Jenny would receive one hour credit. At the same time, Mike would have to give some credit points to Jenny. If he likes Jenny's work, he can give certain number of positive points. Meanwhile, if Mike is not satisfied, he can give certain number of negative points. As the service provider, Jenny can cumulate her credit points through this way. However, this would not be Jenny's final points. Because how Mike gives credit points to Jenny is unknown, there may be a bias in the evaluation procedure. For instance, maybe Mike just does not like the way Jenny talks but Jenny actually has clean the home very well if objectively speaking. Therefore, by collecting huge amount of points from users, an algorithm must be created to calculate the final points of each user to ensure objectiveness, fairness and accuracy. 

\section{Conclusions and Discussions}
\label{sec:conclusion}  
Motivated by the need of a new decentralized social trust infrastructure to serve our community in the era witnessing the proliferation of big data and AI power, this position paper proposes BLESS, a blockchain-enabled social credits system. Due to the decentralization nature of the blockchain networks, data collection and sharing functions are executed by different entities without relying on a centralized third-party authority. The consensus mechanism in blockchain offers tamper proof of transaction data in the trustless network environment. The smart contract based credit rating strategy enables encapsulated rating algorithms. The credit scores become transparent to all participants, and attempts to tamper with the rating process or unauthorized modifying credit scores are prevented. 

The BLESS system is expected to improve the transparency to public events and allow individuals to oversee the operations conducted by those in power, such as big enterprises and organizations. It encourages people to actively participate the social credit system construction, and those who have trust-keeping behaviors will get benefit of more credits and obtain higher priority in the community. Consequently, the BLESS system will encourage the residents in the communities to behave nice and make the entire community more trustworthy and safer.

From technical point of view, the integrated system design and implementation should not be very difficult based on existing technologies. From social science perspective, however, both the social credit system and the time banking face their own challenges in public acceptance and mass adoption. Success of a decentralized social trust infrastructure requires extensive efforts by experts from multiple disciplinary, including social science, political science, public policy, financial and businesses. It is non-trivial to estimate the value of people's efforts in community beneficial activities. 

This position paper by no means implies a matured, comprehensive solution. The authors hope this initial, preliminary idea inspire more discussion and attract attentions of experts from different areas, particularly social science.   


\bibliographystyle{IEEEtranS} 
\bibliography{references} 

\begin{thebibliography}{10}
\providecommand{\url}[1]{#1}
\csname url@samestyle\endcsname
\providecommand{\newblock}{\relax}
\providecommand{\bibinfo}[2]{#2}
\providecommand{\BIBentrySTDinterwordspacing}{\spaceskip=0pt\relax}
\providecommand{\BIBentryALTinterwordstretchfactor}{4}
\providecommand{\BIBentryALTinterwordspacing}{\spaceskip=\fontdimen2\font plus
\BIBentryALTinterwordstretchfactor\fontdimen3\font minus
  \fontdimen4\font\relax}
\providecommand{\BIBforeignlanguage}[2]{{%
\expandafter\ifx\csname l@#1\endcsname\relax
\typeout{** WARNING: IEEEtranS.bst: No hyphenation pattern has been}%
\typeout{** loaded for the language `#1'. Using the pattern for}%
\typeout{** the default language instead.}%
\else
\language=\csname l@#1\endcsname
\fi
#2}}
\providecommand{\BIBdecl}{\relax}
\BIBdecl

\bibitem{TimeBanksUSA}
``{TimeBanks USA},'' \url{http://timebanks.org/}, accessed September 3, 2018.

\bibitem{bellotti2014towards}
V.~M. Bellotti, S.~Cambridge, K.~Hoy, P.~C. Shih, L.~R. Handalian, K.~Han, and
  J.~M. Carroll, ``Towards community-centered support for peer-to-peer service
  exchange: rethinking the timebanking metaphor,'' in \emph{Proceedings of the
  SIGCHI Conference on Human Factors in Computing Systems}.\hskip 1em plus
  0.5em minus 0.4em\relax ACM, 2014, pp. 2975--2984.

\bibitem{cahn2011time}
E.~Cahn, ``Time banking: An idea whose time has come,'' \emph{Yes Magazine},
  2011.

\bibitem{cahn2000no}
E.~S. Cahn, \emph{No more throw-away people: The co-production
  imperative}.\hskip 1em plus 0.5em minus 0.4em\relax Edgar Cahn, 2000.

\bibitem{carroll2013co}
J.~M. Carroll, ``Co-production scenarios for mobile time banking,'' in
  \emph{International Symposium on End User Development}.\hskip 1em plus 0.5em
  minus 0.4em\relax Springer, 2013, pp. 137--152.

\bibitem{chorzempa2018china}
M.~Chorzempa, P.~Triolo, S.~Sacks \emph{et~al.}, ``China’s social credit
  system: A mark of progress or a threat to privacy?'' Tech. Rep., 2018.

\bibitem{drum2017job}
K.~Drum, ``You will lose your job to a robot—and sooner than you think,''
  \emph{\url{https://www.motherjones.com/politics/2017/10/you-will-lose-your-job-to-a-robot-and-sooner-than-you-think/}},
  2017.

\bibitem{hammi2018bubbles}
M.~T. Hammi, B.~Hammi, P.~Bellot, and A.~Serhrouchni, ``Bubbles of trust: A
  decentralized blockchain-based authentication system for iot,''
  \emph{Computers \& Security}, vol.~78, pp. 126--142, 2018.

\bibitem{kwon2017community}
M.~Kwon, C.~Lee, Y.~Xiao, and W.~A. McIntosh, ``Community currency activities,
  community attachment, and quality of life: A case study of the crooked river
  alliance of timebanks,'' \emph{Time \& Society}, p. 0961463X17716737, 2017.

\bibitem{Lee2018ai}
K.-F. Lee, ``The human promise of the ai revolution,'' \emph{The Wall Street
  Journal}, no. September 14, 2018.

\bibitem{liang2018constructing}
F.~Liang, V.~Das, N.~Kostyuk, and M.~M. Hussain, ``Constructing a data-driven
  society: China's social credit system as a state surveillance
  infrastructure,'' \emph{Policy \& Internet}, 2018.

\bibitem{mester1997s}
L.~J. Mester \emph{et~al.}, ``What's the point of credit scoring?''
  \emph{Business review}, vol.~3, no. Sep/Oct, pp. 3--16, 1997.

\bibitem{nagothu2018microservice}
D.~Nagothu, R.~Xu, S.~Y. Nikouei, and Y.~Chen, ``A microservice-enabled
  architecture for smart surveillance using blockchain technology,''
  \emph{arXiv preprint arXiv:1807.07487}, 2018.

\bibitem{nakamoto2008bitcoin}
S.~Nakamoto, ``Bitcoin: A peer-to-peer electronic cash system,'' 2008.

\bibitem{nikouei2018real}
S.~Y. Nikouei, R.~Xu, D.~Nagothu, Y.~Chen, A.~Aved, and E.~Blasch, ``Real-time
  index authentication for event-oriented surveillance video query using
  blockchain,'' \emph{arXiv preprint arXiv:1807.06179}, 2018.

\bibitem{obrien2018cnn}
S.~A. O'Brien, N.~Black, C.~Devine, and D.~Griffin, ``Cnn investigation: 103
  uber drivers accused of sexual assault or abuse,''
  \emph{\url{https://money.cnn.com/2018/04/30/technology/uber-driver-sexual-assault/index.html}},
  2018.

\bibitem{selby2018vaccine}
D.~Selby, ``How a fake vaccine scandal in china is causing a crisis of
  confidence,''
  \emph{\url{https://www.globalcitizen.org/en/content/china-fake-vaccine-rabies-scandal-dpt/}},
  2018.

\bibitem{swan2015blockchain}
M.~Swan, \emph{Blockchain: Blueprint for a new economy}.\hskip 1em plus 0.5em
  minus 0.4em\relax " O'Reilly Media, Inc.", 2015.

\bibitem{szabo1997formalizing}
N.~Szabo, ``Formalizing and securing relationships on public networks,''
  \emph{First Monday}, vol.~2, no.~9, 1997.

\bibitem{wang2018didi}
Y.~Wang and R.~Olsen, ``China's didi chuxing faces intense pressure amid public
  anger over second passenger death,''
  \emph{\url{https://www.forbes.com/sites/ywang/2018/09/01/chinas-didi-chuxing-faces-intense-pressure-amid-public-anger-over-second-passenger-death/\#6d4bf787a6d0}},
  2018.

\bibitem{xu2018blendcac}
R.~Xu., Y.~Chen, E.~Blasch, and G.~Chen, ``Blendcac: A blockchain-enabled
  decentralized capability-based access control for iots,'' in \emph{the IEEE
  International Conference on Blockchain, Selected Areas in IoT and
  Blockchain}.\hskip 1em plus 0.5em minus 0.4em\relax IEEE, 2018.

\bibitem{xu2018smartcac}
R.~Xu, Y.~Chen, E.~Blasch, and G.~Chen, ``Blendcac: A smart contract enabled
  decentralized capability-based access control mechanism for the iot,''
  \emph{Computers 2018, 7(3), 39; Access on:
  http://www.mdpi.com/2073-431X/7/3/39}, 2018.

\bibitem{yuan2018assessing}
C.~W. Yuan, B.~V. Hanrahan, and J.~M. Carroll, ``Assessing timebanking use and
  coordination: implications for service exchange tools,'' \emph{Information
  Technology \& People}, 2018.

\end{thebibliography}

\end{document}